\documentclass[10pt,a4paper]{article}
\usepackage{authblk}
\usepackage[latin1]{inputenc}
\usepackage{amsmath}
\usepackage{amsfonts}
\usepackage{amssymb}
\usepackage{tabularx}
\usepackage{graphicx}
\usepackage{slashed}
\usepackage{subfigure}
\usepackage{subfig}
\usepackage{fullpage}
\begin{document}

\title{Mixing of pseudoscalar-baryon and vector-baryon in the $J^P=1/2^-$ sector and the $N^*$(1535) and $N^*$(1650) resonances.}
\author[1]{E. J. Garzon}
\author[1]{E. Oset}
\affil[1]{Departamento de F\'{\i}sica Te\'orica and IFIC, Centro Mixto Universidad de Valencia-CSIC,
Institutos de Investigaci\'on de Paterna, Aptdo. 22085, 46071 Valencia, Spain}

\date{\today}
\maketitle
\begin{abstract}
We study the meson-baryon interaction with $J^P=1/2^-$ using the hidden-gauge Lagrangians and mixing pseudoscalar meson-baryon with the vector meson-baryon states in a coupled channels scheme with $\pi N$, $\eta N$, $K\Lambda$, $K\Sigma$, $\rho N$ and $\pi \Delta$ (d-wave). We fit the subtraction constants of each channel to the $S_{11}$ partial wave amplitude of the $\pi N$ scattering data extracted from experimental data. We find two poles that we associate to the $N^*(1535)$ and the $N^*(1650)$ resonances and show that the subtraction constants are all negative and of natural size. We calculate the branching ratios for the different channels of each resonance and we find a good agreement with the experimental data. The cross section for the $\pi^- p \to \eta n$ scattering is also evaluated and compared with experiment.
\end{abstract}

\section{Introduction}
Partial wave analyses of $\pi N$ data \cite{Arndt:2006bf,Workman:2012jf} have provided us with 
much data on amplitudes, cross sections and resonance properties. It has also been the subject of intense theoretical investigations (see Refs.~\cite{Doring:2014qaa,Kamano:2013iva} for recent updates on the subject). The introduction of the chiral unitary techniques to study these reactions in Ref.~\cite{Kaiser:1995cy} resulted in surprising news that the $N^*(1535)$ resonance was dynamically generated from the interaction of meson baryon, with a price to pay: coupled channels had to be introduced. Some of the channels were closed at certain energies, like the $K \Lambda$ and $K \Sigma$ in the region of the $N^*(1535)$, but they were shown to play a major role in the generation of this resonance, to the point of suggesting in Ref.~\cite{Kaiser:1995cy} that the $N^*(1535)$ could qualify as a quasibound state of $K \Lambda$ and $K \Sigma$. Work on this issue followed in Ref.~\cite{Inoue:2001ip}, corroborating the main findings of Ref.~\cite{Kaiser:1995cy}, and posteriorly in Refs.~\cite{Hyodo:2002pk,
Kolomeitsev:2003kt,GarciaRecio:2003ks,Hyodo:2008xr}. In the chiral unitary approach the loops of the Bethe-Salpeter equation must be regularized, and this is done with cut offs or using dimensional regularization. The cut off, or equivalently the subtraction constants in dimensional regularization in the different channels should be of ``natural size'', as discussed in Ref.~\cite{Hyodo:2008xr,Oller:2000fj,Sekihara:2014kya}, if one wishes to claim that the resonances have been generated dynamically from the interaction. However, this is not the case of the $N^*(1535)$, where different cut offs in Ref.~\cite{Kaiser:1995cy}, or different subtraction constants in Ref.~\cite{Inoue:2001ip} for different channels must be used. This is unlike the case of the $\Lambda(1405)$, where a unique cut off in all channels leads to a good reproduction of the data \cite{Oset:1997it,Oller:2000fj,GarciaRecio:2002td,Jido:2003cb}. This fact was interpreted in Ref.~\cite{Hyodo:2008xr} as a manifestation of the nature of the two resonances, where the $\Lambda(1405)$ would be largely dynamically generated, while the $N^*(1535)$ would contain a nonnegligible component of a genuine state, formed with dynamics different from the pseudoscalar meson interaction. One might think of remnants of an original seed of three constituent quarks, but this is not necessarily the case. It could also be due to the missing of important channels different than pseudoscalar-baryon. Actually this has been a source of investigation recently, where the mixing of pseudoscalar-baryon and vector-baryon channels has led to interesting results and some surprises. In Ref.~\cite{Garzon:2012np} the vector-baryon interaction was studied using the method developed in Ref.~\cite{Oset:2009vf} but mixing also pseudoscalar-baryon components. It was found that the mixing produced a shift of some of the resonance positions of Ref.~\cite{Oset:2009vf} and led to some increase in the width. Similar results have been obtained recently in Refs.~\cite{Khemchandani:2011et,Khemchandani:2011mf,Khemchandani:2012ur}. One of the interesting outcomes of this line of research was to see that the consideration of the $\pi N$(d-wave), $\rho N$, $\pi \Delta$(s and d waves) in the sector of spin-parity $3/2^-$ with chiral dynamics led to a good reproduction of the $\pi N$ data in d-waves and to the generation of the $N^*(1520)$ and $N^*(1700)$ resonances \cite{Garzon:2013pad}.
In Ref.~\cite{Nieves:2001wt} $N^*(1535)$ and $N^*(1650)$ are obtained with only pseudoscalar - baryon states using an offshell approach, which is in principle equivalent to having different subtraction constants in different channels. By means of that one can effectively take into account the effect of missing channels. In our work we explicitly introduce from the beginning a larger space of channels.
 In the present work we want to extend the results of Ref.~\cite{Garzon:2013pad} to the sector of $1/2^-$, with the aim to see if the mixture of the pseudoscalar-baryon and vector-baryon channels can remove the pathology observed by the need of different subtraction constants in different channels. We will show that this is the case and then we shall be able to conclude that the missing components of the wave function in the $N^*(1535)$ noted in Ref.~\cite{Hyodo:2008xr} are due to vector-baryon and additional $\pi \Delta$ states that we shall also mix in the present coupled channel approach.

\section{Formalism}
\begin{figure}
\begin{center}
\includegraphics[scale=0.4]{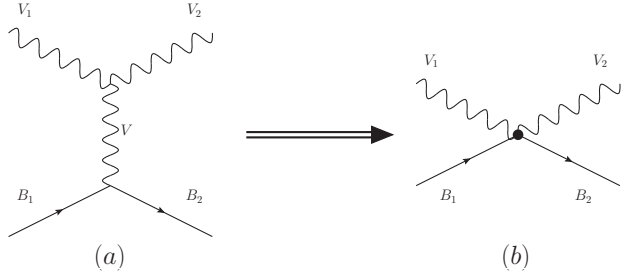}
\caption[Vector meson-baryon interaction.]{Vector meson-baryon interaction: (a) with vector meson exchange (b) contact term.}
\label{fig:VBcontact}
\end{center}
\end{figure}
The most important coupled channels of $N^*(1535)$ and $N^*(1650)$ are $\pi N$, $\eta N$, $K\Lambda$, $K\Sigma$, $\rho N$ and $\pi \Delta$ (d-wave). Some of the matrix elements of the interaction between these channels have been well studied as the $PB\to PB$ transition mediated by a vector meson exchange addressed in Ref.~\cite{Inoue:2001ip}. The diagram involved in this transition is show in Fig.~\ref{fig:VBcontact} and the potential of this transition is given by
\begin{equation}
V_{i j}= - C_{i j} \, \frac{1}{4 f^2} \, \left( k^0 + k^\prime{}^0\right)
\end{equation}
The $PB$ transition coefficients are taken from Ref.~\cite{Inoue:2001ip}. However, since those coefficients are in charge basis we need to convert them to isospin basis, as show in Table~\ref{tab:coeffPB}. Similarly, the $\rho N \to \rho N$ transition has been studied in Ref.~\cite{Garzon:2013pad} and the coefficients are given in Appendix A of Ref.~\cite{Oset:2009vf}.
\begin{table}
\begin{center}
\setlength{\tabcolsep}{3.0pt}
\renewcommand*\arraystretch{1.5}
\begin{tabular}{c|cccc}
&$\pi N$&$\eta N$&$K\Lambda$&$K\Sigma$\\
\hline
$\pi N$&2&0&$\frac{3}{2}$&$-\frac{1}{2}$\\
$\eta N$&&0&$-\frac{3}{2}$&$-\frac{3}{2}$\\
$K\Lambda$&&&0&0\\
$K\Sigma$&&&&2\\
\end{tabular}
\end{center}
\caption{Coefficients of $PB$ transition with $I=1/2$}
\label{tab:coeffPB}
\end{table}
\begin{figure}
\begin{center}
\subfigure[$\rho N(s) \rightarrow \rho N(s)$]{
\includegraphics[width=0.2\textwidth]{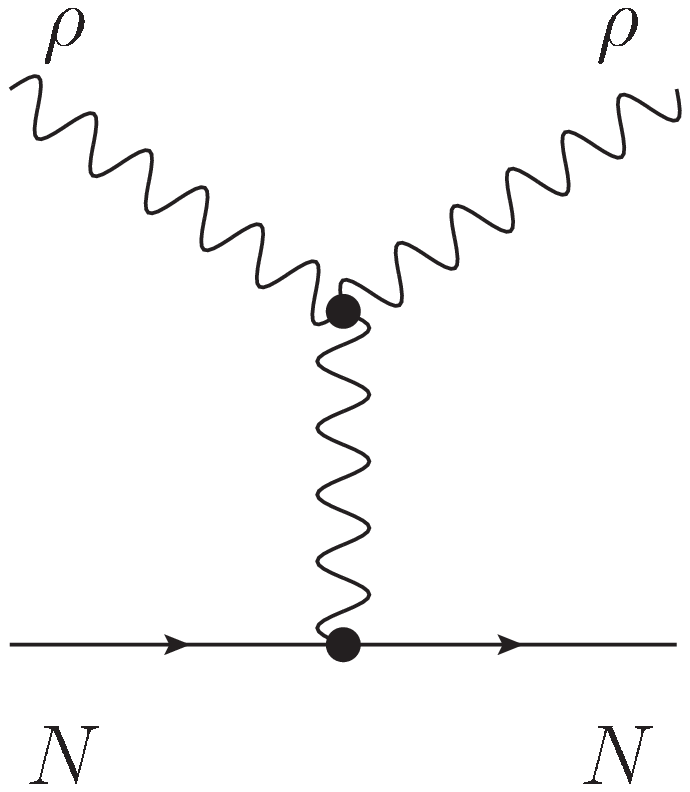}}
\subfigure[$\pi \Delta(s) \rightarrow \pi \Delta(s)$]{
\includegraphics[width=0.2\textwidth]{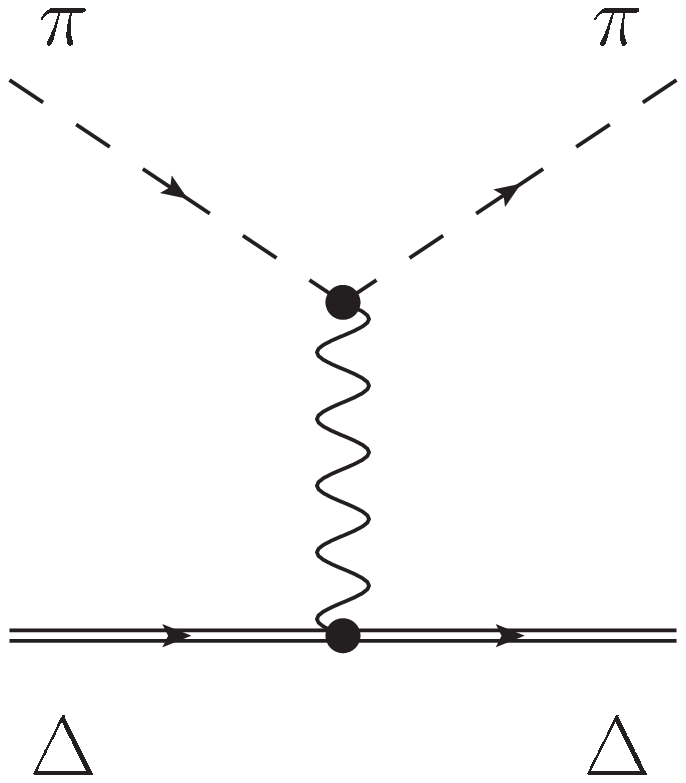}}
\subfigure[$\rho N(s) \rightarrow \pi N(d)$]{
\includegraphics[width=0.2\textwidth]{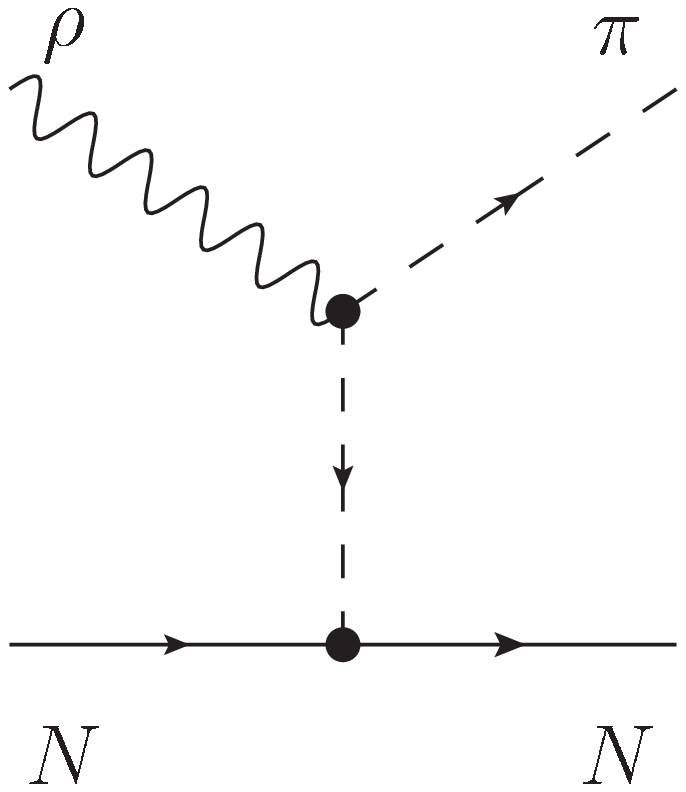}}
\\
\subfigure[$\rho N(s) \rightarrow \pi \Delta(s,d)$]{
\includegraphics[width=0.2\textwidth]{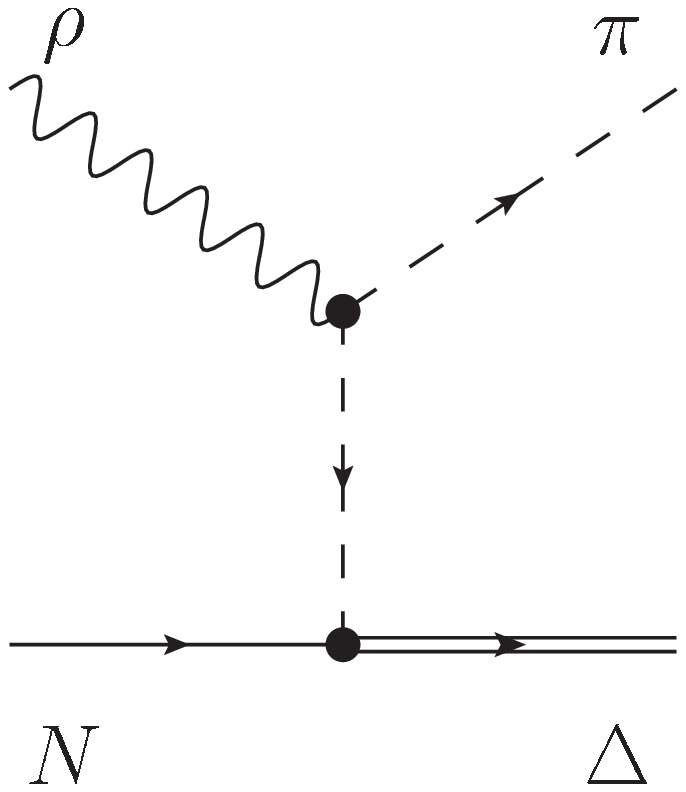}}
\subfigure[$\rho N(s) \rightarrow \pi \Delta(s)$]{
\includegraphics[width=0.2\textwidth]{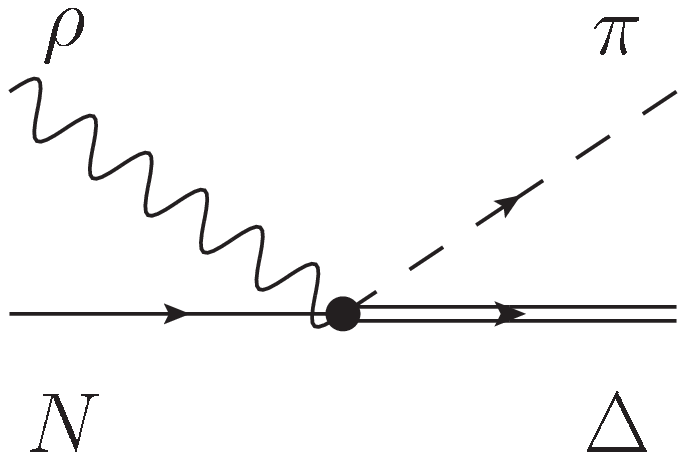}}
\caption{Diagrams of the channels involved in the calculation for $N^*(1520)$ and $N^*(1700)$.}
\label{fig:diagrams}
\end{center}
\end{figure}
The transition to $V B\to PB$ is implemented following the formalism described in Ref.~\cite{Garzon:2013pad}, where the interaction is mediated by a pseudoscalar meson as shown in Fig.~\ref{fig:diagrams}(c). Furthermore we also include the Kroll-Ruderman term shown in Fig.~\ref{fig:diagrams}(e). The evaluation of the diagrams shown in Fig.~\ref{fig:diagrams} leads us to the following vertices for the transitions (See Appendix A of Ref.~\cite{Garzon:2013pad} for details)
\begin{eqnarray}
t_{\rho N (s) \rightarrow \pi N (s)}&=	&- 2\sqrt{6} g \frac{D+F}{2 f} \left\lbrace \frac{\frac{2}{3} \vec{q}\,_{\pi N}^2}{\left(P_V+q_{\pi N}\right)^2-m_\pi^2}+1 \right\rbrace \\
t_{\rho N (s) \rightarrow \eta N (s)}&=&0 \\
t_{\rho N (s) \rightarrow K \Lambda (s)}&=&-\frac{1}{2}\sqrt{6} g \frac{D+3F}{2 f} \left\lbrace \frac{\frac{2}{3} \vec{q}\,_{K \Lambda}^2}{\left(P_V+q_{K \Lambda}\right)^2-m_K^2}+1 \right\rbrace \\
t_{\rho N (s) \rightarrow K \Sigma (s)}&=&-\frac{1}{2}\sqrt{6} g \frac{D-F}{2 f} \left\lbrace \frac{\frac{2}{3} \vec{q}\,_{K \Sigma}^2}{\left(P_V+q_{K \Sigma}\right)^2-m_K^2}+1 \right\rbrace \\
\end{eqnarray}
where we take $F=0.51$ and $D=0.75$ \cite{Borasoy:1998pe,Close:1993mv}, an $q_i$ is the momentum of the pseudoscalar meson in the center of mass. The factor $1$ that appears inside the braces corresponds to the Kroll-Ruderman vertex. In comparison with the results of Ref.~\cite{Khemchandani:2013nma}, where the authors only take into account the Kroll-Rudermann term, we obtain the same coefficients for the $PB\to VB$ transition.

\begin{figure}
\begin{center}
\includegraphics[scale=0.5]{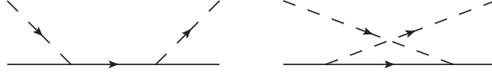}
\end{center}
\caption{Diagrams of $s-$ and $u-$channels exchange with the nucleon propagator.}
\label{fig:suchannel}
\end{figure}

Moreover we find interesting to include the contribution to the s-wave from the $s-$ and $u-$channels containing the nucleon propagator, shown in Fig.~\ref{fig:suchannel}, given by
\begin{equation}
t_{\pi N \to N \to \pi N}= (E_\pi)^2 \left( \frac{D+F}{2 f} \right)^2 \left(\frac{3}{\sqrt{s}+M_N}- 
\frac{1}{\sqrt{s}-2E_\pi +M_N} \right) 
\end{equation}
\begin{equation}
t_{\eta N \to N \to \eta N}= (E_\eta)^2 \left(\frac{1}{\sqrt{3}} \frac{D-3F}{2 f} \right)^2 \left(\frac{1}{\sqrt{s}+M_N}+ 
\frac{1}{\sqrt{s}-2E_\eta +M_N} \right) 
\end{equation}
This is easily obtained by separating the relativistic nucleon propagator into positive and negative energy components 
\begin{equation}
\frac{\slashed p+m}{p^2-m^2}=\frac{M}{E(\vec{p}\,)} \sum_r \left\lbrace 
\frac{u_r(\vec{p}\,) \bar{u}_r(\vec{p}\,)}{p^0-E(\vec{p}\,)+i\epsilon}+
\frac{ v_r(-\vec{p}\,) \bar{v}_r(-\vec{p}\,)}{p^0+E(\vec{p}\,)-i\epsilon} \right\rbrace
\end{equation}
Then the positive energy part contributes to p-wave and the negative energy part to s-wave. We should note that these terms, as well as a possible isoscalar seagull contribution \cite{Meissner:1999vr,Doring:2004kt} give a very small contribution.

On the other hand we have the transition of $\rho N \to \pi \Delta (d)$ that has been already studied in Ref.~\cite{Garzon:2013pad}. The diagram of this transition is given in Fig.~\ref{fig:diagrams}(d) and the evaluation of this diagram gives the transition given by
\begin{equation}
t_{\rho N (s) \rightarrow \pi \Delta (d)}=g \frac{2}{\sqrt{3}} \frac{f_{\pi N \Delta}}{m_\pi} \left\lbrace \frac{\frac{2}{3} \vec{q}\,^2}{\left(P_V+q\right)^2-m_\pi^2}\right\rbrace
\end{equation}
Here we do not have the $1$ factor from the Kroll-Ruderman term since this transition only involves $L=2$.

As done in Ref.~\cite{Garzon:2013pad} the transitions involving $L=2$ are introduced with a parameter $\gamma$. This is done for the diagonal transition $\pi \Delta (d) \to \pi \Delta (d)$ and for the transition channel that we consider relevant $\pi \Delta (d) \to \pi N (s)$.
\begin{equation}
t_{\pi \Delta (d) \to \pi \Delta (d)} = -\frac{\gamma_0}{m_{\pi}^5} q_{\pi \Delta}^4
\end{equation}
\begin{equation}
t_{\pi \Delta (d) \to \pi N (s)} = -\frac{\gamma_{1}}{m_{\pi}^3} q_{\pi \Delta}^2
\end{equation}
The parameters are normalized with the corresponding power of the pion mass to be dimensionless. Both are parametrized, but since the $\pi \Delta (d) \to \pi \Delta (d)$ transition are both d-wave channels, the potential has four momenta while the transition $\pi \Delta (d) \to \pi N (s)$ is a transition from s-wave to d-wave has only two momenta. As done before, the divergence of the momenta is controlled with the Blatt-Weisskopf barrier-penetration factors as done in Ref.~\cite{Garzon:2013pad}.
 
\section{Fitting the data}
\begin{figure}
\begin{center}
\includegraphics[scale=0.5]{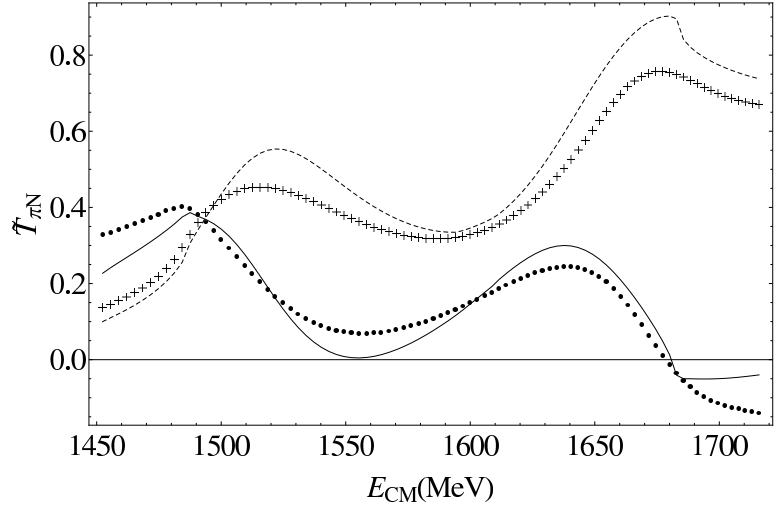}
\caption{Fit to the data extracted from Ref.~\cite{Arndt:1995bj}. We show the real part (circles) and imaginary part (crosses) of the data and the result of our fit of $\tilde{T}_{\pi N}$ for the real (solid) and imaginary (dashed) parts.}
\label{fig:fitN1535}
\end{center}
\end{figure}

We have some unknown parameters in our theory that we need to determine fitting the data. First we have the subtraction constant for each channel which, are expected to be around $-2$ with a regularization scale of $\mu =630$ MeV. We have also two undetermined parameters in the potential $\gamma_0$ and $\gamma_1$ corresponding to the transition of $\pi \Delta (d) \to \pi \Delta (d)$ and $\pi \Delta (d) \to \pi N$. We perform a fit of the $S_{11}$ partial wave amplitude of the $\pi N$ scattering data extracted from experimental data of Ref.~\cite{Arndt:1995bj}. We need to normalize the amplitude of the T-matrix using Eq.~(7) of Ref.~\cite{Roca:2006sz}, which relates our amplitude with the experimental one by
\begin{equation}
\tilde{T}_{ij}(\sqrt{s})=-\sqrt{\frac{M_i q_i}{4\pi \sqrt{s}}}\sqrt{\frac{M_j q_j}{4\pi \sqrt{s}}} T_{ij}(\sqrt{s})
\end{equation}
where $M$ is the mass of the baryon for the specific channel and $q$ is the on-shell momentum. In Fig.~\ref{fig:fitN1535} we show the fit of both the real and imaginary parts of $\tilde{T}$ for the diagonal channel of $\pi N$. 

In Table \ref{tab:parfitN1535} we show the results obtained with the fit. As we commented before, we have used a regularization scale different for each channel, corresponding to the mass of the baryon of that channel. We consider interesting to show also the subtraction constant with a regularization scale of $\mu=630$ MeV in order to compare them with other results found in the literature. It is interesting to see that the introduction of the $\rho N$ and $\pi \Delta$ channels has had an important qualitative effect in the subtraction constants, which now are all negative and of the same order of magnitude, while in Ref.~\cite{Inoue:2001ip} some of the subtraction constants were even positive.

\begin{table}
\begin{center}
\setlength{\tabcolsep}{5.0pt}
\renewcommand*\arraystretch{1.5}	
\begin{tabular}{ccccccccc}
\hline
\hline
$\mu$[MeV]& $a_{N\pi}$& $a_{N\eta}$& $a_{\Lambda K}$& $a_{\Sigma K}$ & $a_{N\rho}$& $a_{\Delta\pi}$& $\gamma_0$&$\gamma_1$\\
\hline
$M_B$&-1.203&-2.208&-1.985&-0.528&-0.493&-1.379&0.595&1.47\\
$630$&-2.001&-3.006&-3.128&-1.799&-1.291&-2.720&0.595&1.47\\
\hline
\hline
\end{tabular}
\caption{Parameters obtained with the fit. The first row are the parameters with a regularization scale $\mu$ that corresponds to the mass of the baryon of each channel. Second row are the same results but with the natural regularization scale $\mu=630$ MeV. The parameters $\gamma_i$ are not changed.}
\label{tab:parfitN1535}
\end{center}
\end{table}

\section{Results}
\begin{figure}[p]
\begin{center}
\includegraphics[scale=0.6]{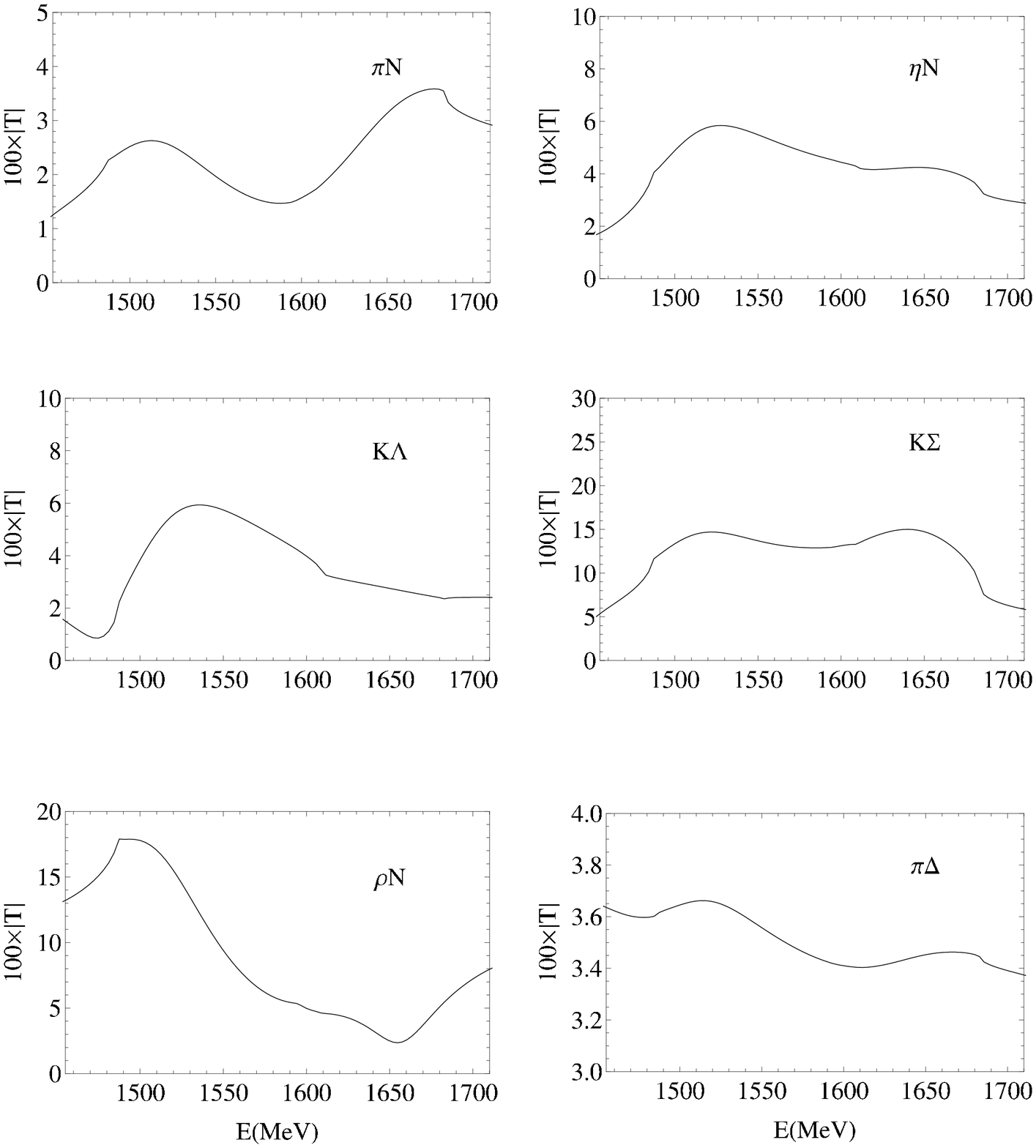}
\caption{Results of the $|T|^2$ matrix for the diagonal channels.}
\label{fig:resN1535}
\end{center}
\end{figure}
Using the parameters determined with the fit, we evaluate the T matrix using the Bethe-Salpeter equation and show in Fig.~\ref{fig:resN1535} the result of $|T|^2$ for all the diagonal channels. 
Analysing the T matrix, using the method explained in detail in Ref.~\cite{Garzon:2013pad}, we found two poles that can be associated to the resonances $N^*(1535)$ and $N^*(1650)$. This is a remarkable novelty, since in Ref.~\cite{Inoue:2001ip} the $N^*(1535)$ appears but not the $N^*(1650)$. The poles are used to calculate the couplings of all channels to each resonances. These results are compiled in Table \ref{tab:couplingsN1535}. With the couplings one can determine the decay width and branching ratio to each channel of both resonances. The results of the branching ratios for the resonances $N^*(1535)$ and $N^*(1650)$ are shown in Tables~\ref{tab:resN1535} and \ref{tab:resN1650} respectively. In the Tables we show the position of the poles and the branching ratios for each channel found in this work. We compare them with the experimental results of the PDG \cite{Agashe:2014kda}, and as the PDG average has big uncertainties we also compare them with single results of the experiments and analysis \cite{Cutkosky:1979fy,Anisovich:2011fc,Vrana:1999nt,Thoma:2007bm}.

\begin{table}
\begin{center}
\begin{small}
\setlength{\tabcolsep}{2.0pt}
\renewcommand*\arraystretch{1.5}
\begin{tabular}{c|rrlcrrl|rrlcrrl}
\hline
\hline
&\multicolumn{7}{c|}{$N^*(1535)$}&\multicolumn{7}{c}{$N^*(1650)$}\\
\hline
Channel&
\multicolumn{3}{c}{$g_i$}&$|g_i|$&\multicolumn{3}{c|}{$g_i G_i$}&
\multicolumn{3}{c}{$g_i$}&$|g_i|$&\multicolumn{3}{c}{$g_i G_i$}\\
\hline
$N\pi$			&1.03&+i&0.21&1.05&- 6.68&-i&24.29&1.37&+i&0.54&1.47&  2.52&-i&36.51\\
$N\eta$			&1.40&+i&0.78&1.60&-30.50&-i&29.20&1.08&-i&0.60&1.24&-33.89&-i&2.51\\
$\Lambda K$		&1.71&+i&0.48&1.78&-38.06&-i&14.50&0.10&-i&0.68&0.69&- 9.96&+i&17.67\\
$\Sigma K$		&1.70&+i&1.24&2.10&  1.58&-i& 2.77&3.21&-i&1.34&3.47&-28.75&-i&13.14\\
$N\rho$			&2.96&+i&0.11&2.96& 17.71&-i& 2.61&0.94&+i&1.51&1.78&  7.83&-i&2.25\\
$\Delta\pi$		&0.31&-i&0.04&0.31&- 8.17&-i& 3.20&0.31&+i&0.03&0.31&- 6.03&-i&6.72\\
\hline
\hline
\end{tabular}
\caption{Couplings of the different channels to each resonance}
\label{tab:couplingsN1535}
\end{small}
\end{center}
\end{table}

Looking at Table~\ref{tab:couplingsN1535} we see that the $\rho N$ channel has the strongest coupling to $N^*(1535)$ but the resonance is below the threshold, however due to the width of the $\rho$ we can generate enough phase space and obtain a small width, but in a very good agreement with the experimental results, as seen in Table~\ref{tab:resN1535}. The $K \Sigma$ channel has a coupling as big as the $\rho N$ but as the resonance is below the threshold it has not phase space to decay. Similarly the coupling to channel $K \Lambda$ is big but again there is not phase space for decay. On the other hand the $\pi \Delta$ channel has a very small coupling but around $200$ MeV of phase space, so this gives it a small branching ratio which agrees with the experimental values. The other channels $\pi N$ and $\eta N$ have smaller couplings but since they have much momentum to decay they have big branching ratios in good agreement with the experimental results. Concerning the width of the $N^*(1535)$ in Table~\ref{tab:resN1535} we should note that, although the theoretical width obtained from the pole in the complex plane is smaller than the experimental one, the apparent width from Im$\tilde{T}_{\pi N}$ in the real axis, seen in Fig.~\ref{fig:fitN1535}, is much closer to the experiment.

For the case of $N^*(1650)$ the $K \Sigma$ channel has now the biggest coupling to this resonance, but as the resonance is below the threshold it has no phase space for decay. The same as before happens to the $\rho N$ channel, the small momentum generated with the mass convolution of the $\rho$ gives a small width but in qualitative agreement with the experimental value. Although the channels $\pi N$, $\eta N$ and $\pi \Delta$ have smaller couplings, due to the huge phase space that they have, the branching ratios are quite big, which is in a very good agreement with the experimental results of PDG, but with single experiments as well, as seen in Table~\ref{tab:resN1650}. Now, the $K \Lambda$ channel is open and the value found for the branching ratio is in good agreement with the only experimental value available of Ref.~\cite{Anisovich:2011fc}.

\begin{table}
\begin{center}
\begin{small}
\setlength{\tabcolsep}{2.0pt}
\renewcommand*\arraystretch{1.5}
\begin{tabular}{r|c|rcl|rcl|rcl|rcl|rcl}
\hline
\hline
\multicolumn{17}{c}{$N^*(1535)~J^P=1/2^-$}\\
\hline
&Theory&
\multicolumn{3}{c|}{PDG}&
\multicolumn{3}{c|}{Cutkosky}&
\multicolumn{3}{c|}{Anisovich}&
\multicolumn{3}{c|}{Vrana}&
\multicolumn{3}{c}{Thoma}\\
&&
\multicolumn{3}{c|}{}&
\multicolumn{3}{c|}{\cite{Cutkosky:1979fy}}&
\multicolumn{3}{c|}{\cite{Anisovich:2011fc}}&
\multicolumn{3}{c|}{\cite{Vrana:1999nt}}&
\multicolumn{3}{c}{\cite{Thoma:2007bm}}\\
\hline
Re(Pole)&
1508.1&1490&--&1530&1510&$\pm$&50&1501&$\pm$&4&&1525&&1508&${}^{+}_{-}$&${}^{10}_{30}$\\
2Im(Pole)&
90.3&90&--&250&260&$\pm$&80&134&$\pm$&11&&102&&165&$\pm$&15\\
\hline
Channel&\multicolumn{16}{c}{Branching Ratio [$\Gamma_i/\Gamma$($\%$)]}\\
\hline
N$\pi$(1077)&
58.6&35&--&55&50&$\pm$&10&54&$\pm$&5&35&$\pm$&8&37&$\pm$&9\\
N$\eta$(1487)&
37.0&42&$\pm$&10&&&&33&$\pm$&5&51&$\pm$&5&40&$\pm$&10\\
$\Lambda$K(1609)&
0.0&&-&&&&&&&&&&&&&\\
$\Sigma$K(1683)&
0.0&&-&&&&&&&&&&&&&\\
N$\rho$(1714)&
1.0&2&$\pm$&1&&&&&&&2&$\pm$&1&&&\\
$\Delta\pi$(1370)&
3.3&0&--&4&&&&2.5&$\pm$&1.5&1&$\pm$&1&23&$\pm$&8\\
\hline
\hline
\end{tabular}
\end{small}
\caption{Results for the pole position and branching ratios for the different channels of $N^*(1535)~J^P=1/2^-$ and comparison with experimental results.}
\label{tab:resN1535}
\end{center}
\end{table}

We consider interesting to include in Table~\ref{tab:couplingsN1535} the value of the wave function in coordinate space at the origin, defined in Ref.~\cite{Gamermann:2009uq} as
\begin{equation}
\left(2\pi \right)^{3/2} \psi_i (\vec{0})=g_i G_i(z_R)
\end{equation}
where the $G$ function is evaluated in the pole. This magnitude represents the wave function at the origin for s-wave channels. For d-wave channels the wave function goes as $r^2$ at the origin and vanishes. The magnitude $gG$ then represents the relative strength of the channel for coupling of the resonance to external sources \cite{YamagataSekihara:2010pj}. The results show information about how relevant is each channel for the resonances. The first surprise is to see that, although the $K\Sigma$ channel has the second biggest coupling, the value of the wave function at the origin reveals that this channel is not relevant in the $N^*(1535)$. We also see that the most important channels are the $\eta N$ and $\pi N$, as one can expect of the experimental results of the branching rations. Moreover the $K \Lambda$ channel has an important contribution but since the resonance is below the threshold, this fact is not noticeable experimentally. For the $N^*(1650)$ case the $\pi N$ channel is now the most important but the $\eta N$ channel is very important as well. However we can see that now the $K\Sigma$ has a very important contribution since the pole is very close to the $K \Sigma$ threshold. The $K \Lambda$ has a moderate relevance and this is in agreement with the experimental results for the width, as seen in Table~\ref{tab:resN1650}. The $\rho N$ and $\pi \Delta$ channels have a small relative contribution and this is in good agreement with experimental values.
\begin{table}
\begin{center}
\begin{small}
\setlength{\tabcolsep}{2.0pt}
\renewcommand*\arraystretch{1.5}
\begin{tabular}{r|c|rcl|rcl|rcl|rcl|rcl}
\hline
\hline
\multicolumn{17}{c}{$N^*(1650)~J^P=1/2^-$}\\
\hline
&Theory&
\multicolumn{3}{c|}{PDG}&
\multicolumn{3}{c|}{Cutkosky}&
\multicolumn{3}{c|}{Anisovich}&
\multicolumn{3}{c|}{Vrana}&
\multicolumn{3}{c}{Thoma}\\
&&
\multicolumn{3}{c|}{}&
\multicolumn{3}{c|}{\cite{Cutkosky:1979fy}}&
\multicolumn{3}{c|}{\cite{Anisovich:2011fc}}&
\multicolumn{3}{c|}{\cite{Vrana:1999nt}}&
\multicolumn{3}{c}{\cite{Thoma:2007bm}}\\
\hline
\hline
Re(Pole)&
1672.3&1640&--&1670&1640&$\pm$&20&1647&$\pm$&6&&1663&&1645&$\pm$&15\\
2Im(Pole)&
158.2&100&--&170&150&$\pm$&30&103&$\pm$&8&&240&&187&$\pm$&20\\
\hline
Channel&\multicolumn{16}{c}{Branching Ratio [$\Gamma_i/\Gamma$($\%$)]}\\
\hline
N$\pi$&
58.9&50&--&90&65&$\pm$&10&51&$\pm$&4&74&$\pm$&2&70&$\pm$&15\\
N$\eta$&
27.6&5&--&15&&&&18&$\pm$&4&6&$\pm$&1&15&$\pm$&6\\
$\Lambda$K&
5.7&&-&&&&&10&$\pm$&5&&&&&&\\
$\Sigma$K&
0.0&&-&&&&&&&&&&&&&\\
N$\rho$&
5.6&1&$\pm$&1&&&&&&&1&$\pm$&1&&&\\
$\Delta\pi$&
2.2&0&--&25&&&&19&$\pm$&9&2&$\pm$&1&10&$\pm$&5\\
\hline
\hline
\end{tabular}
\end{small}
\caption{Results for the pole position and branching ratios for the different channels of $N^*(1650)~J^P=1/2^-$ and comparison with experimental results.}
\label{tab:resN1650}
\end{center}
\end{table}

\begin{figure}
\begin{center}
\includegraphics[scale=0.5]{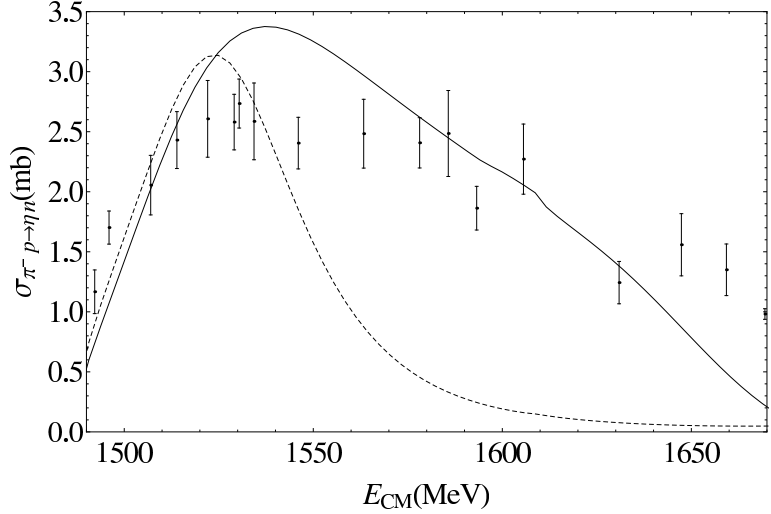}
\end{center}
\caption{Comparison of the cross section of $\pi^- p \to \eta n$ scattering (solid) and the result for the cross section of Ref.~\cite{Inoue:2001ip} (dashed).}
\label{fig:cspieta}
\end{figure}

One reaction that filters the $I=1/2$ and $J^P=1/2^-$, that we study here, is the $\pi^- p \to \eta n$. In Fig.~\ref{fig:cspieta} we show the result for the cross section of the $\pi^- p \to \eta n$, where we can see a good agreement with the experimental data of Ref.~\cite{Baldini:1988th}. In comparison with the results of Ref.~\cite{Inoue:2001ip} for the same reaction, we see a good improvement of our work, since in Ref.~\cite{Inoue:2001ip} the experimental data of the cross section above 1550 MeV is not well reproduced. In this work we generate dynamically the $N^*(1650)$ resonance, which fills up the region of the cross section above 1550 MeV that is not well obtained in Ref.~\cite{Inoue:2001ip}, where the $N^*(1650)$ was not generated.

\section{Conclusions}
We have studied the meson-baryon interaction with $J^P=1/2^-$ including the coupled channels considered in the experimental analysis, $\pi N$, $\eta N$, $K\Lambda$, $K\Sigma$, $\rho N$ and $\pi \Delta$ (d-wave). We have studied the interaction using the hidden gauge formalism, where the interaction is mediated by the exchange of vector mesons. Other extensions of this formalism involving pseudoscalar and vector mesons are also used as explained in the text. The loops are regularized using dimensional regularization with subtraction constants for each channel. These constants are treated as free parameters and fitted to reproduce the experimental data of the $S_{11}$ $\pi N$ scattering data extracted from Ref.~\cite{Arndt:1995bj}.

Two poles are found and the couplings for each channel, as well as the wave function at the origin, are calculated. These couplings are used to obtain the branching ratios to all channels of both resonances. The results are then compared with several experimental values and there is a good agreement for most of them. It must be noted that the consideration of the $\rho N$ and $\pi \Delta$ channels has had an important qualitative change with respect to the work of Ref.~\cite{Inoue:2001ip} where only the pseudoscalar-baryon octet channels were considered. The first one is that now we are able to generate both the $N^*(1535)$ and the $N^*(1650)$ resonances, while in Ref.~\cite{Inoue:2001ip} only the $N^*(1535)$ appeared. The second one is that now the subtraction constants are all negative and of natural size. From the perspective of Ref.~\cite{Hyodo:2008xr} we can say that the conclusion in Ref.~\cite{Hyodo:2008xr} that the $N^*(1535)$ had an important component of a genuine state in the wave function, can be translated now by stating that the missing components can be filled up by the $\rho N$ and $\pi \Delta$ channels that we have found here.

\end{document}